\documentclass[a4paper]{jpconf}

\usepackage{graphicx}
\usepackage{amsmath}
\usepackage{xcolor}
\usepackage{hyperref}

\begin{document}

\title{An open-source modular framework for quantum computing}

\author{S. Carrazza$^{1,2,3}$, S. Efthymiou$^3$, M. Lazzarin$^3$, A. Pasquale$^{1,3}$}

\address{$^1$ Dipartimento di Fisica, Universit\`a degli Studi di Milano and INFN Sezione di Milano.}
\address{$^2$ CERN, Theoretical Physics Department, CH-1211 Geneva 23, Switzerland.}
\address{$^3$ Quantum Research Center, Technology Innovation Institute, Abu Dhabi, UAE.}

\ead{stefano.carrazza@cern.ch}

\begin{abstract}
In this proceedings we describe the current development status and recent
technical achievements of Qibo, an open-source framework for quantum simulation.
After a concise overview of the project goal, we introduce the modular
layout for backend abstraction released in version 0.1.7. We discuss the
advantages of each backend choice with particular emphasis on hardware
accelerators for quantum state vector simulation. Finally, we summarize the
primitives and models currently available.
\end{abstract}

\section{Introduction}

Quantum computing is a new paradigm whereby quantum phenomena are harnessed to
perform computations. The current availability of noisy intermediate-scale
quantum (NISQ) computers ~\cite{nisq}, combined with recent advances towards quantum
computational supremacy~\cite{supremacy, zhong2020quantum}, has led to a
growing interest in these devices to perform computational tasks faster than
classical machines. Among many of the near-term
applications~\cite{cerezo2021variational, bharti2021noisy}, the field of Quantum
Machine Learning (QML)~\cite{biamonte2017quantum, schuld2018supervised} is held
as one promising approach to make use of NISQ computers, including applications
to evolving research fields such as High-Energy
Physics~\cite{qpdf2021,bravoprieto2021stylebased}.

Nowadays, quantum processing units (QPUs) are based on two major approaches.
The first one is based on quantum circuit and quantum logic gate-based model processors,
as implemented most popularly by
Google~\cite{google}, IBM~\cite{ibmq}, Rigetti~\cite{rigetti} or
Intel~\cite{intel}. The second employs annealing quantum processors such as
D-Wave~\cite{dwave,dwaveneal} among others. The development of these devices and the
achievement of quantum advantage~\cite{48651} are indicators that a
technological revolution in computing will occur in the coming years.
However, in parallel to the development of QPU technology, we still have to
perform classical simulation of quantum computing, which has been at the
cornerstone of quantum research, to elaborate new algorithms and applications.
From a theoretical perspective it serves as the basic tool for testing and
developing quantum algorithms, while from an experimental point of view it
provides a platform for benchmarks and error simulation.

Circuit-based quantum computers can be classically simulated using
Schr\"odinger's or Feynman's
approach~\cite{boixo2017simulation,chen2018classical}. The former is based on
keeping track of the full quantum state and applying gates via specialized matrix
multiplication routines. The latter, inspired by Feynman's path integrals, can
be used to calculate amplitudes of the final state by summing over different
histories (paths). Schr\"odinger's approach is memory intensive as it requires
storing the full quantum state consisting of $2^n$ complex numbers for $n$ qubits,
however its run time is linear on the number of gates in the circuit. Feynman's
approach memory requirements scale linearly with the number of qubits and gates,
however its run time grows exponentially with the number of gates~\cite{bvcomplexity, aaronsoncomplexity}.
Hybrid methods exploiting both approaches to achieve a run time vs performance
trade-off have also been explored~\cite{feynmanhybrid}.

\begin{figure}
    \centering
    \includegraphics[width=0.85\textwidth]{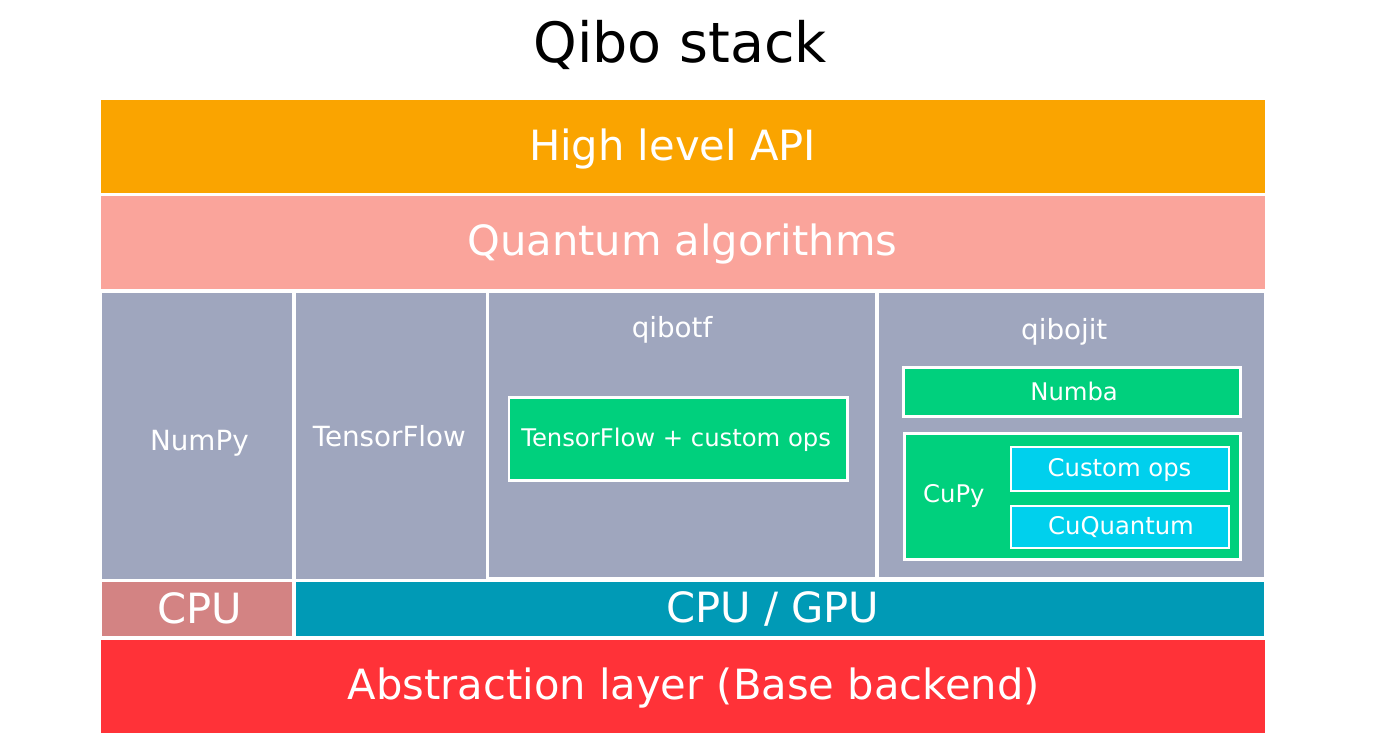}
    \caption{Schematic view of structure design in Qibo version 0.1.7.}
    \label{fig:stack}
\end{figure}

Qibo~\cite{qibo2021,stavros_efthymiou_2021_5711842} is an open-source framework
for quantum computing which supports general purpose quantum simulation based on
Schr\"odinger's approach.
The source code of Qibo, available at \url{https://github.com/qiboteam/qibo}, provides a
high-level API for writing quantum circuits and gates, abstraction layers for
simulation and hardware control backends and a collection of pre-coded quantum
algorithms and research inspired examples.
Its structure is visualized in Fig.~\ref{fig:stack} and will be discussed in the
next sections using as reference the latest release version 0.1.7.

\section{Quantum simulation backends}
\label{sec:backends}

A system of $n$ qubits is described by a state vector $\psi $ of $2^n$ complex
numbers, which represent the probability amplitudes in the computational basis.
A gate targeting $n_\mathrm{tar}$ qubits can be represented as a
$2^{n_\mathrm{tar}} \times 2^{n_\mathrm{tar}}$ complex matrix $G$.
In Schr\"odinger's approach of quantum simulation, each gate is applied to the
state via the following matrix multiplication
\begin{equation}\label{eq:gateapplication}
    \psi'(\boldsymbol{\tau}, \boldsymbol{q}) =
    \sum _{\boldsymbol{\tau'}}
    G(\boldsymbol{\tau},\boldsymbol{\tau'})
    \psi(\boldsymbol{\tau'}, \boldsymbol{q})
\end{equation}
where $\boldsymbol{\tau}$ and $\boldsymbol{q}$ denote bitstrings of length
$n_\mathrm{tar}$ and $n - n_\mathrm{tar}$ respectively and the sum runs
over all possibile bitstrings $\boldsymbol{\tau'}$ of length $n_\mathrm{tar}$.
From a computational point of view, such paradigm opens the possibility to use
different techniques and hardware to achieve efficient simulation of the final state.

The Qibo package, distributed from PyPI\footnote{\url{https://pypi.org/project/qibo/}}
and conda-forge\footnote{\url{https://anaconda.org/conda-forge/qibo}},
is shipped with two basic simulators (\texttt{numpy} and \texttt{tensorflow})
which can efficiently simulate circuits of up to 20 qubits.
This base package provides an abstraction
layer written in Python which defines an abstract backend class with a minimal
set of methods for linear algebra manipulation and gate application.
The {\tt numpy} simulator is based on NumPy~\cite{numpybook} primitives
supporting only single thread CPU, while the {\tt tensorflow} simulator replaces
these primitives with those of TensorFlow~\cite{tensorflow2015-whitepaper}. These choices
allow the simple execution of quantum circuits on multi-threading CPU and GPU
configurations.
The multiplications in Eq.~\ref{eq:gateapplication} are implemented using
the \texttt{numpy.einsum} and \texttt{tensorflow.einsum} methods that provide
a generic algorithm with moderate performance.
Despite the performance limitation, the \texttt{numpy} backend is relevant for
cross-platform deployment. NumPy supports a high number of architectures,
including \texttt{arm64}, allowing the \texttt{numpy} backend to be deployed in
several contexts, such as laboratories developing QPUs, where servers
do not always match the \texttt{x86\_64} architecture.
In addition, the \texttt{tensorflow} backend provides automatic gradient
evaluation for gradient descent optimization. This combination allows the
development of variational quantum circuits for quantum machine learning. This
choice opens the possibility to develop novel hybrid classical-quantum models
such as quantum generative adversarial
networks~\cite{bravoprieto2021stylebased}.

Additional backends are available as add-on packages and provide higher
performance for larger circuits. The {\tt qibotf}
package\footnote{\url{https://github.com/qiboteam/qibotf}} was the first high
performance backend included in the first Qibo release (0.1.0). It is based on
TensorFlow custom operators written in C++ and CUDA. In contrast to TensorFlow
primitives, custom operators perform in-place updates, \textit{i.e.}, the state is not duplicated
during circuit execution. This reduces both memory requirements and execution time.
The main disadvantages associated with {\tt qibotf} are the need to maintain C++ code,
and the compilation of custom operators before execution which slows down
development and complicates installation by reducing the target of potential
devices that could benefit from pre-compiled binaries.

To address these issues, we developed the {\tt qibojit}
backend~\cite{stavros_efthymiou_2021_5248470}
package\footnote{\url{https://github.com/qiboteam/qibojit}} which supports
execution on multi-threading CPU, GPU, and multi-GPU configurations.
The CPU part of {\tt qibojit} uses custom operators like {\tt qibotf},
which are now written in Python and compiled just-in-time using Numba~\cite{lam2015numba}.
The loop over the state elements is parallelized using OpenMP~\cite{openmp} via
Numba's \texttt{numba.prange} method. Each element is updated according to
the rule defined in Eq.~\ref{eq:gateapplication}.
The proper indices for each update, which depend on target qubit index,
are generated on-the-fly during the loop using fast binary operations.
Gates that target multiple qubits and controlled gates can be
applied similarly after modifying the index generation accordingly.
The GPU part uses CUDA kernels that follow the same approach as CPU but are
written in C++ and compiled just-in-time using CuPy~\cite{cupy_learningsys2017}.
The kernels are exposed to Python using CuPy's RawModule.
Compatibility with CuPy also allowed us to incorporate to \texttt{qibojit} the recently
released quantum simulation library cuQuantum~\cite{cuquantum} by NVIDIA.
Exploiting Numba and Cupy capabilities simplifies the code without sacrificing performance.
It also makes the installation on different platforms easier.
Exhaustive performance benchmarks between the just-in-time and pre-compiled
approaches for quantum simulation together with a technical explanation of
techniques used to accellerate simulation performance are addressed in Sections
II and III in Ref.~\cite{Efthymiou:2022apj}.

\section{Primitives and models}

\begin{figure}
    \centering
    \includegraphics[width=0.8\textwidth]{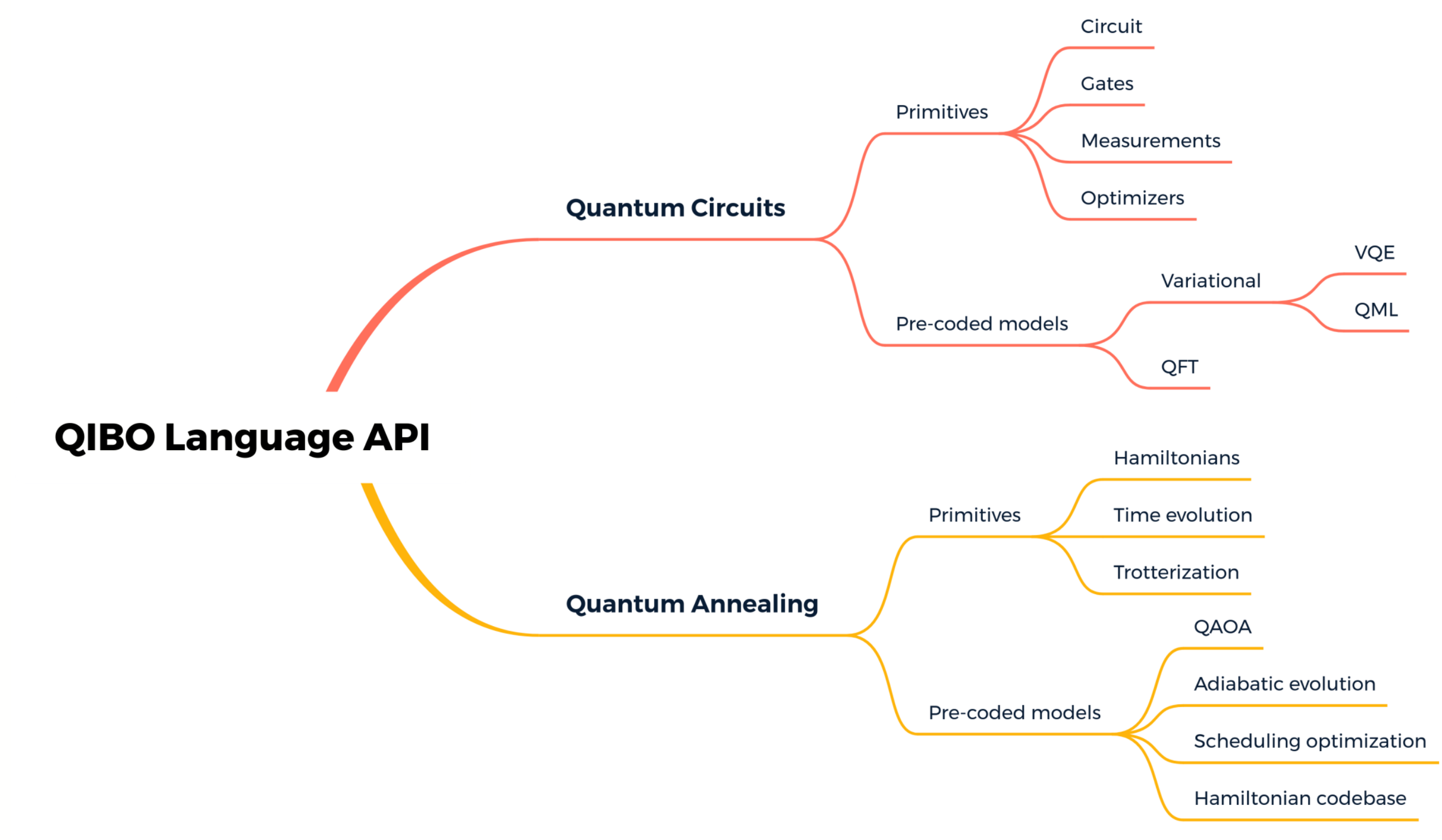}
    \caption{Models and primitives in Qibo version 0.1.7.}
    \label{fig:models}
\end{figure}

In Figure~\ref{fig:models} we show the current components available when
installing the base Qibo 0.1.7 package. At the current stage, the framework
supports quantum simulation using circuits and annealing paradigms, which are
most likely the two representations that can be deployed on real QPU devices.

The quantum circuit representation is composed of a set of primitives which
allow the user to allocate quantum circuits with gates
based on single-, two-, and three-qubits. The simulation of measurements is
provided through special sampling kernels for each backend. We have integrated
classical optimization algorithms based on the approximate Hessian
approach~\cite{2020SciPy-NMeth}, evolutionary
strategies~\cite{nikolaus_hansen_2021_5002422} and gradient descent optimization
through TensorFlow.
In terms of built-in models, we provide variational gate layers which can be
efficiently trained using the optimizers mentioned above. Furthermore, we
include circuits such as the quantum Fourier transform (QFT), Grover's algorithm
and the variational quantum eigensolver (VQE) for fast implementation and
benchmark. Finally, we provide models for quantum machine learning, generative
adversarial networks (style-qGAN)~\cite{bravoprieto2021stylebased}, quantum
regressors~\cite{qpdf2021} and several tutorials involving quantum
classifiers~\cite{classifier2020}.

Concerning the quantum annealing paradigm, we provide primitives for Hamiltonian
representation, for both numerical and analytic representations. These objects are
interfaced with a time evolution solver and the possibility to use the Trotter
decomposition, as presented in Sec.~4.1 of~\cite{Paeckel:2019yjf}. We provide a
large codebase of pre-coded Hamiltonians, such as non-interacting Pauli-X/Y/Z,
transverse field Ising model, MaxCut, and Heisenberg XXZ. Furthermore, we
include pre-coded models for adiabatic evolution~\cite{farhi2000quantum}, with
the possibility to determine the best parametric scheduling function, and
adiabatically assisted variational quantum eigensolver
(AAVQE)~\cite{garciasaez2018addressing}, quantum approximate optimization
algorithms (QAOA)~\cite{farhi2014quantum}, and the feedback-based algorithm for
quantum optimization (FALQON)~\cite{magann2021feedbackbased}.

\section{Outlook}

The latest Qibo release (0.1.7) includes a modular approach to quantum
simulation engines with support on multi-threading CPU, GPU, and multi-GPU
hardware setups. We provide backends for multiple architectures, including those
which are popular in experimental laboratories developing QPU technologies.
The primitives and models implemented in the code are in continuous expansion
and are driven by the requirements and feedback from users developing spin-off
projects involving quantum computing.

In the future we are planning to explore the implementation of alternative
approaches to state vector simulation (Schr\"odinger's approach).
Furthermore, we will start testing the framework on real quantum hardware by
expanding the current portfolio of backends to experimental setups.

\section*{References}
\bibliographystyle{elsarticle-num}
\bibliography{references}

\begin{thebibliography}{10}
\expandafter\ifx\csname url\endcsname\relax
  \def\url#1{\texttt{#1}}\fi
\expandafter\ifx\csname urlprefix\endcsname\relax\def\urlprefix{URL }\fi
\expandafter\ifx\csname href\endcsname\relax
  \def\href#1#2{#2} \def\path#1{#1}\fi

\bibitem{nisq}
J.~Preskill, \href{http://dx.doi.org/10.22331/q-2018-08-06-79}{Quantum
  computing in the {NISQ} era and beyond}, Quantum 2 (2018) 79.
\newblock \href {https://doi.org/10.22331/q-2018-08-06-79}
  {\path{doi:10.22331/q-2018-08-06-79}}.
\newline\urlprefix\url{http://dx.doi.org/10.22331/q-2018-08-06-79}

\bibitem{supremacy}
F.~Arute, K.~Arya, R.~Babbush, D.~Bacon, J.~C. Bardin, R.~Barends, R.~Biswas,
  S.~Boixo, F.~G.~S. L.~Brandao, D.~A. Buell, et~al.,
  \href{http://doi.org/10.1038/s41586-019-1666-5}{Quantum supremacy using a
  programmable superconducting processor}, Nature 574~(7779) (2019) 505--510.
\newblock \href {https://doi.org/10.1038/s41586-019-1666-5}
  {\path{doi:10.1038/s41586-019-1666-5}}.
\newline\urlprefix\url{http://doi.org/10.1038/s41586-019-1666-5}

\bibitem{zhong2020quantum}
H.-S. Zhong, H.~Wang, Y.-H. Deng, M.-C. Chen, L.-C. Peng, Y.-H. Luo, J.~Qin,
  D.~Wu, X.~Ding, Y.~Hu, et~al., Quantum computational advantage using photons,
  Science 370~(6523) (2020) 1460--1463.

\bibitem{cerezo2021variational}
M.~Cerezo, A.~Arrasmith, R.~Babbush, S.~C. Benjamin, S.~Endo, K.~Fujii, J.~R.
  McClean, K.~Mitarai, X.~Yuan, L.~Cincio, et~al., Variational quantum
  algorithms, Nature Reviews Physics 3 (2021) 625--644.

\bibitem{bharti2021noisy}
K.~Bharti, A.~Cervera-Lierta, T.~H. Kyaw, T.~Haug, S.~Alperin-Lea, A.~Anand,
  M.~Degroote, H.~Heimonen, J.~S. Kottmann, T.~Menke, et~al., Noisy
  intermediate-scale quantum ({NISQ}) algorithms, arXiv preprint
  arXiv:2101.08448 (2021).

\bibitem{biamonte2017quantum}
J.~Biamonte, P.~Wittek, N.~Pancotti, P.~Rebentrost, N.~Wiebe, S.~Lloyd, Quantum
  machine learning, Nature 549~(7671) (2017) 195--202.

\bibitem{schuld2018supervised}
M.~Schuld, F.~Petruccione, Supervised learning with quantum computers, Vol.~17,
  Springer, 2018.

\bibitem{qpdf2021}
A.~Pérez-Salinas, J.~Cruz-Martinez, A.~A. Alhajri, S.~Carrazza,
  \href{http://dx.doi.org/10.1103/PhysRevD.103.034027}{Determining the proton
  content with a quantum computer}, Physical Review D 103~(3) (Feb 2021).
\newblock \href {https://doi.org/10.1103/physrevd.103.034027}
  {\path{doi:10.1103/physrevd.103.034027}}.
\newline\urlprefix\url{http://dx.doi.org/10.1103/PhysRevD.103.034027}

\bibitem{bravoprieto2021stylebased}
C.~Bravo-Prieto, J.~Baglio, M.~Cè, A.~Francis, D.~M. Grabowska, S.~Carrazza,
  Style-based quantum generative adversarial networks for {Monte Carlo} events
  (2021).
\newblock \href {http://arxiv.org/abs/2110.06933} {\path{arXiv:2110.06933}}.

\bibitem{google}
{Google Research},
  \href{https://research.google/teams/applied-science/quantum/}{{Google AI
  Quantum}} (2017).
\newline\urlprefix\url{https://research.google/teams/applied-science/quantum/}

\bibitem{ibmq}
{IBM Research}, \href{https://www.ibm.com/quantum-computing/}{{IBM Quantum
  Experience}} (2016).
\newline\urlprefix\url{https://www.ibm.com/quantum-computing/}

\bibitem{rigetti}
{Rigetti}, \href{https://www.rigetti.com/}{{Rigetti Computing}} (2017).
\newline\urlprefix\url{https://www.rigetti.com/}

\bibitem{intel}
{Intel Corporation},
  \href{https://www.intel.com/content/www/us/en/research/quantum-computing.html}{{Intel
  Quantum Computing}} (2017).
\newline\urlprefix\url{https://www.intel.com/content/www/us/en/research/quantum-computing.html}

\bibitem{dwave}
{D-Wave Systems}, \href{https://www.dwavesys.com/}{{The Quantum Computing
  Company}} (2011).
\newline\urlprefix\url{https://www.dwavesys.com/}

\bibitem{dwaveneal}
{D-Wave Systems}, \href{https://github.com/dwavesystems/dwave-neal}{{D-Wave
  Neal}}.
\newline\urlprefix\url{https://github.com/dwavesystems/dwave-neal}

\bibitem{48651}
F.~A. et~al., Quantum supremacy using a programmable superconducting processor,
  Nature 574 (2019) pp. 505--510.
\newblock \href {https://doi.org/10.1038/s41586-019-1666-5}
  {\path{doi:10.1038/s41586-019-1666-5}}.

\bibitem{boixo2017simulation}
S.~Boixo, S.~V. Isakov, V.~N. Smelyanskiy, H.~Neven, Simulation of low-depth
  quantum circuits as complex undirected graphical models (2017).
\newblock \href {http://arxiv.org/abs/1712.05384} {\path{arXiv:1712.05384}}.

\bibitem{chen2018classical}
J.~Chen, \textit{et al.}, Classical simulation of intermediate-size quantum
  circuits (2018).
\newblock \href {http://arxiv.org/abs/1805.01450} {\path{arXiv:1805.01450}}.

\bibitem{bvcomplexity}
E.~Bernstein, U.~Vazirani,
  \href{https://doi.org/10.1137/S0097539796300921}{Quantum complexity theory},
  SIAM Journal on Computing 26~(5) (1997) 1411--1473.
\newblock \href
  {http://arxiv.org/abs/https://doi.org/10.1137/S0097539796300921}
  {\path{arXiv:https://doi.org/10.1137/S0097539796300921}}, \href
  {https://doi.org/10.1137/S0097539796300921}
  {\path{doi:10.1137/S0097539796300921}}.
\newline\urlprefix\url{https://doi.org/10.1137/S0097539796300921}

\bibitem{aaronsoncomplexity}
S.~Aaronson, L.~Chen, Complexity-theoretic foundations of quantum supremacy
  experiments, in: Proceedings of the 32nd Computational Complexity Conference,
  CCC '17, Schloss Dagstuhl--Leibniz-Zentrum fuer Informatik, Dagstuhl, DEU,
  2017.

\bibitem{feynmanhybrid}
I.~L. Markov, A.~Fatima, S.~V. Isakov, S.~Boixo, Quantum supremacy is both
  closer and farther than it appears, arXiv preprint arXiv:1807.10749 (2018).

\bibitem{qibo2021}
S.~Efthymiou, S.~Ramos-Calderer, C.~Bravo-Prieto, A.~P{\'{e}}rez-Salinas,
  D.~Garc{\'{\i}}a-Mart{\'{\i}}n, A.~Garcia-Saez, J.~I. Latorre, S.~Carrazza,
  \href{https://doi.org/10.1088/2058-9565/ac39f5}{Qibo: a framework for quantum
  simulation with hardware acceleration}, Quantum Science and Technology 7~(1)
  (2021) 015018.
\newblock \href {https://doi.org/10.1088/2058-9565/ac39f5}
  {\path{doi:10.1088/2058-9565/ac39f5}}.
\newline\urlprefix\url{https://doi.org/10.1088/2058-9565/ac39f5}

\bibitem{stavros_efthymiou_2021_5711842}
{The Qibo team}, \href{https://doi.org/10.5281/zenodo.3997194}{qiboteam/qibo:
  Qibo}.
\newblock \href {https://doi.org/10.5281/zenodo.3997194}
  {\path{doi:10.5281/zenodo.3997194}}.
\newline\urlprefix\url{https://doi.org/10.5281/zenodo.3997194}

\bibitem{numpybook}
T.~Oliphant, Guide to NumPy, 2006.

\bibitem{tensorflow2015-whitepaper}
M.~Abadi, A.~Agarwal, P.~Barham, E.~Brevdo, Z.~Chen, C.~Citro, G.~S. Corrado,
  A.~Davis, J.~Dean, M.~Devin, S.~Ghemawat, I.~Goodfellow, A.~Harp, G.~Irving,
  M.~Isard, Y.~Jia, R.~Jozefowicz, L.~Kaiser, M.~Kudlur, J.~Levenberg,
  D.~Man\'{e}, R.~Monga, S.~Moore, D.~Murray, C.~Olah, M.~Schuster, J.~Shlens,
  B.~Steiner, I.~Sutskever, K.~Talwar, P.~Tucker, V.~Vanhoucke, V.~Vasudevan,
  F.~Vi\'{e}gas, O.~Vinyals, P.~Warden, M.~Wattenberg, M.~Wicke, Y.~Yu,
  X.~Zheng, \href{https://www.tensorflow.org/}{{TensorFlow}: Large-scale
  machine learning on heterogeneous systems}, software available from
  tensorflow.org (2015).
\newline\urlprefix\url{https://www.tensorflow.org/}

\bibitem{stavros_efthymiou_2021_5248470}
S.~Efthymiou, S.~Carrazza,
  \href{https://doi.org/10.5281/zenodo.5071354}{qiboteam/qibojit: qibojit}.
\newblock \href {https://doi.org/10.5281/zenodo.5071354}
  {\path{doi:10.5281/zenodo.5071354}}.
\newline\urlprefix\url{https://doi.org/10.5281/zenodo.5071354}

\bibitem{lam2015numba}
S.~K. Lam, A.~Pitrou, S.~Seibert, Numba: A llvm-based python jit compiler, in:
  Proceedings of the Second Workshop on the LLVM Compiler Infrastructure in
  HPC, 2015, pp. 1--6.

\bibitem{openmp}
{The OpenMP development team}, \href{https://www.openmp.org/}{{OpenMP
  website}}.
\newline\urlprefix\url{https://www.openmp.org/}

\bibitem{cupy_learningsys2017}
R.~Okuta, Y.~Unno, D.~Nishino, S.~Hido, C.~Loomis,
  \href{http://learningsys.org/nips17/assets/papers/paper_16.pdf}{{CuPy}: A
  {NumPy}-compatible library for {NVIDIA} {GPU} calculations}, in: Proceedings
  of Workshop on Machine Learning Systems (LearningSys) in The Thirty-first
  Annual Conference on Neural Information Processing Systems (NIPS), 2017.
\newline\urlprefix\url{http://learningsys.org/nips17/assets/papers/paper_16.pdf}

\bibitem{cuquantum}
NVIDIA, \href{https://developer.nvidia.com/cuquantum-sdk}{{cuQuantum SDK}}
  (2021).
\newline\urlprefix\url{https://developer.nvidia.com/cuquantum-sdk}

\bibitem{Efthymiou:2022apj}
S.~Efthymiou, M.~Lazzarin, A.~Pasquale, S.~Carrazza, {Quantum simulation with
  just-in-time compilation} (3 2022).
\newblock \href {http://arxiv.org/abs/2203.08826} {\path{arXiv:2203.08826}}.

\bibitem{2020SciPy-NMeth}
P.~Virtanen, R.~Gommers, T.~E. Oliphant, M.~Haberland, T.~Reddy, D.~Cournapeau,
  E.~Burovski, P.~Peterson, W.~Weckesser, J.~Bright, S.~J. {van der Walt},
  M.~Brett, J.~Wilson, K.~J. Millman, N.~Mayorov, A.~R.~J. Nelson, E.~Jones,
  R.~Kern, E.~Larson, C.~J. Carey, {\.I}.~Polat, Y.~Feng, E.~W. Moore,
  J.~{VanderPlas}, D.~Laxalde, J.~Perktold, R.~Cimrman, I.~Henriksen, E.~A.
  Quintero, C.~R. Harris, A.~M. Archibald, A.~H. Ribeiro, F.~Pedregosa, P.~{van
  Mulbregt}, {SciPy 1.0 Contributors}, {SciPy} 1.0: Fundamental algorithms for
  scientific computing in {Python}, Nature Methods 17 (2020) 261--272.
\newblock \href {https://doi.org/10.1038/s41592-019-0686-2}
  {\path{doi:10.1038/s41592-019-0686-2}}.

\bibitem{nikolaus_hansen_2021_5002422}
N.~Hansen, yoshihikoueno, ARF1, K.~Nozawa, M.~Chan, Y.~Akimoto, D.~Brockhoff,
  \href{https://doi.org/10.5281/zenodo.5002422}{Cma-es/pycma} (Jun. 2021).
\newblock \href {https://doi.org/10.5281/zenodo.5002422}
  {\path{doi:10.5281/zenodo.5002422}}.
\newline\urlprefix\url{https://doi.org/10.5281/zenodo.5002422}

\bibitem{classifier2020}
A.~Pérez-Salinas, A.~Cervera-Lierta, E.~Gil-Fuster, J.~I. Latorre,
  \href{http://dx.doi.org/10.22331/q-2020-02-06-226}{Data re-uploading for a
  universal quantum classifier}, Quantum 4 (2020) 226.
\newblock \href {https://doi.org/10.22331/q-2020-02-06-226}
  {\path{doi:10.22331/q-2020-02-06-226}}.
\newline\urlprefix\url{http://dx.doi.org/10.22331/q-2020-02-06-226}

\bibitem{Paeckel:2019yjf}
S.~Paeckel, \textit{et al.}, Time-evolution methods for matrix-product states,
  Annals of Physics 411 (2019) pp. 167998.
\newblock \href {https://doi.org/10.1016/j.aop.2019.167998}
  {\path{doi:10.1016/j.aop.2019.167998}}.

\bibitem{farhi2000quantum}
E.~Farhi, J.~Goldstone, S.~Gutmann, M.~Sipser, Quantum computation by adiabatic
  evolution (2000).
\newblock \href {http://arxiv.org/abs/quant-ph/0001106}
  {\path{arXiv:quant-ph/0001106}}.

\bibitem{garciasaez2018addressing}
A.~Garcia-Saez, J.~I. Latorre, Addressing hard classical problems with
  adiabatically assisted variational quantum eigensolvers (2018).
\newblock \href {http://arxiv.org/abs/1806.02287} {\path{arXiv:1806.02287}}.

\bibitem{farhi2014quantum}
E.~Farhi, J.~Goldstone, S.~Gutmann, A quantum approximate optimization
  algorithm (2014).
\newblock \href {http://arxiv.org/abs/1411.4028} {\path{arXiv:1411.4028}}.

\bibitem{magann2021feedbackbased}
A.~B. Magann, K.~M. Rudinger, M.~D. Grace, M.~Sarovar, Feedback-based quantum
  optimization (2021).
\newblock \href {http://arxiv.org/abs/2103.08619} {\path{arXiv:2103.08619}}.

\end{thebibliography}

\end{document}